# Quantum behavior of the dc SQUID phase qubit

Kaushik Mitra, F. W. Strauch, C. J. Lobb, J. R. Anderson, F. C. Wellstood

Center for Nanophysics and Advanced Materials and Joint Quantum Institute

Department of Physics, University of Maryland, College Park, Maryland 20742

and

Eite Tiesinga

Joint Quantum Institute, National Institute of Standards and Technology

Gaithersburg, Maryland 20899

**Abstract**

We analyze the behavior of a dc Superconducting Quantum Interference Device (SQUID) phase qubit in which one junction acts as a phase qubit and the rest of the device provides isolation from dissipation and noise in the bias leads.  Ignoring dissipation, we find the two-dimensional Hamiltonian of the system and use numerical methods and a cubic approximation to solve Schrödinger's equation for the eigenstates, energy levels, tunneling rates, and expectation value of the currents in the junctions. Using these results, we investigate how well this design provides isolation while preserving the characteristics of a phase qubit. In addition, we show that the expectation value of current flowing through the isolation junction depends on the state of the qubit and can be used for non-destructive read out of the qubit state.

## I. INTRODUCTION

There are currently several types of superconducting devices that are being actively investigated for use as quantum bits. [1-10] These devices can be grouped into three broad classes: charge, flux and phase qubits, according to which dynamical variable is most sharply defined and which basis states are used. In this paper we examine the quantum behavior of the dc Superconducting Quantum Interference Device (SQUID) phase qubit, investigate the optimization of the device and discuss how well this design provides isolation while preserving the characteristics of a phase qubit. This type of phase qubit [1, 2, 10, 11] was first proposed by Martinis *et al.* [2] and has two junctions in a superconducting loop, just as in a conventional dc SQUID [12,13].

The quantum behavior of dc SQUIDs has been of interest for nearly three decades. Much early research was driven by the desire to construct quantum limited amplifiers and magnetic sensors [14] as well as understanding intrinsic quantum mechanical effects in the SQUID [15]. However, most of this prior work on amplifiers and sensors involved SQUIDs that were shunted with resistors and biased into a finite-voltage power-dissipating state to achieve a non-hysteretic response. In contrast, for use as a qubit, one needs to understand the behavior of SQUID's that are in the low-dissipation limit (no resistive shunt across the junctions) and the device is biased in the zero-voltage state. In particular, we need to understand how to choose the device parameters and arrange the bias so that the device is well-isolated from its leads.

The dc SQUID phase qubit is one type of phase qubit. The archetypal or ideal phase qubit is just a single Josephson tunnel junction connected to a current bias source

[see Fig. 1(a)]. The behavior of an ideal phase qubit is analogous to that of a ball trapped in a tilted washboard potential [see Fig. 1(b)], where the position is proportional to the phase difference across the junction, the size of the ripples in the potential is proportional to the junction's critical current, the tilt is proportional to the applied current, and the mass of the ball is proportional to the junction's capacitance [16]. Quantization of the system yields discrete metastable energy levels as well as a continuum of levels [16]. If the junction capacitance and critical current are sufficiently large, and the tilt not too large, the metastable levels have classically a well-defined phase (hence the name "phase qubit"), corresponding to the ball being trapped in one well of the potential and the lossless flow of supercurrent through the tunnel junction. The two lowest meta-stable energy levels in a well can be used as qubit states ($|0>$ and $|1>$) and their separation in energy can be tuned by changing the bias current. The continuum states correspond to the ball rolling down the potential and a voltage being developed across the junction. The qubit states can decay via quantum tunneling to the continuum, with $|1>$ escaping two or three orders of magnitude faster than $|0>$, and monitoring the decay rate allows sensitive detection of the state.

The phase qubit has some potentially significant advantages, including simplicity, the ability to tune the transition frequency and a fast built-in read-out capability. However, there are also significant problems with this simple design:

(1) While the ability to tune the transition frequency by applying a bias current is very useful, fluctuations in the bias current will cause technical noise like fluctuations in the transition frequency, leading to dephasing and inhomogenous broadening.

(2) Wire leads are needed to connect the bias current source, which is typically at room temperature, to the junction, which must be cooled to milli-Kelvin temperatures. These leads will present a dissipative impedance to the junction, decreasing lifetime $T_1$ for the excited state and a producing correspondingly short coherence time.

(3) While tunneling to the voltage state can provide a fast, high-fidelity readout of the state, it is a highly intrusive measurement that not only causes the qubit state to leave the computational basis, but also causes the dissipation of relatively large amounts of energy at the junction.

(4) Dielectric loss and two-level charge fluctuators in the tunnel barrier, the substrate or nearby insulation layers, or critical current fluctuators in the tunnel junction barrier, can all lead to significant dissipation and dephasing, limiting the coherence time of the device [17].

The main idea behind the design of the dc SQUID phase qubit [2] is to overcome the first two problems by inserting a broad-band inductive isolation network between the qubit junction and the bias leads. In Reference [2] it was demonstrated that measurement of the qubit states can be achieved with fidelity of 85%. However, so far no quantitative analysis has been done illustrating how well the device is protected from bias current noise and on the influence of the multilevel nature of the qubit. Here we will give both a classical and quantum derivation of the device examining these issues.

Overcoming the third problem requires implementing a state readout technique, such as microwave reflectometry, that does not require tunneling to the voltage state.

Here we will examine another possible approach to non-destructive read-out in this device. Overcoming the fourth problem will require the use of low-loss materials with a low density of defects, especially in the tunnel barrier [17].

Other approaches are possible. For example, by operating at a sweet spot, where the transition frequency is independent of current, one can minimize decoherence from current fluctuations. In the phase qubit this requires biasing the device at zero current and choosing the device parameters to maintain sufficient anharmonicity. This is the approach taken in the transmon [18]. Such an approach could be used in conjunction with broad-band isolation. However, since removing the ability to apply bias current will sacrifice tunabilty, one needs to consider the trade-off involved.

Figure 1(c) shows a circuit schematic of the dc SQUID phase qubit. In this device, junction $J1$ behaves like a phase qubit and the rest of the circuit is used to provide isolation from dissipation and low-frequency bias-current noise. The lowest two energy levels of junction $J1$, labeled $|0\rangle$ and $|1\rangle$, are used as the qubit states. An external current source provides a bias current $I$ to the circuit and a second current source supplies current $I_f$ to mutual inductance $M$ in order to apply a flux $\Phi_a = MI_f$ to the SQUID loop. The mutual inductance $M$ must be small enough to ensure that too much noise does not couple to the SQUID via this coil. The second junction J2 is called the isolation junction and is used to provide a state readout capability, for example via tunneling to the voltage state.

To protect junction $J1$ from fluctuations in the bias current $I$, the inductances $L_1$ and $L_2$ are chosen so that $L_1 \gg L_2$; $L_1$ and $L_2$ act like an inductive current divider. To bias $J1$ and set the spacing between energy levels, both currents $I$ and $I_f$ are applied.

Since $L_1 \gg L_2$, when $I$ is applied most of the current goes through junction *J2*, leaving *J1* unbiased. If positive flux $\Phi_a$ is then applied to the SQUID, a circulating current is induced that opposes the initial applied flux. If just the right magnitude of flux is applied, this produces a current flow through *J2* that exactly cancels the current due to the bias current *I*, and leaves current flowing just in *J1*. In this situation, the effective current flowing through the qubit junction *J1* is $I$ while any fluctuation in the bias current mainly flows through the isolation junction *J2*.

In the next section, we discuss the SQUID Hamiltonian and potential energy for typical bias conditions. Our analysis treats both junctions quantum mechanically and ignores dissipation. In Section III, we develop a cubic approximation for the potential and in Section IV we discuss numerical solutions of Schrodinger's equation for the exact Hamiltonian. In Section V we compare the energy levels found using these two approaches and conclude that the results are very nearly the same under typical bias conditions. In Section VI, we analyze how well the qubit junction is isolated from current noise and compare our results to classical analysis of the circuit. In Section VII, we identify novel features in the coupling between the state of the qubit and the circulating current, and discuss the implications of this for non-destructive state readout. Finally, we conclude with a brief summary.

## II. THE CIRCUIT HAMILTONIAN AND EFFECTIVE POTENTIAL

We consider the dc SQUID circuit shown in Fig. 1(c) and show that its dynamics is described by the Lagrangian:

$$\mathcal{L}(\gamma_1, \dot{\gamma}_1, \gamma_2, \dot{\gamma}_2) = \frac{1}{2} m_1 \dot{\gamma}_1^2 + \frac{1}{2} m_2 \dot{\gamma}_2^2 - U(\gamma_1, \gamma_2) \qquad (1)$$

where $m_i = (\Phi_0/2\pi)^2 C_i$ is the effective mass of the $i$-th junction, $\Phi_0 = h/2e$ is the flux quantum, $C_i$ is the capacitance of the i-th junction, and $\gamma_1$ and $\gamma_2$ are the guage invariant phase differences across junctions $J1$ and $J2$. The potential energy term U is:

$$U(\gamma_1,\gamma_2) = -E_{J1}\cos\gamma_1 - \frac{\Phi_0}{2\pi}I\frac{L_2}{L}\gamma_1 - E_{J2}\cos\gamma_2 - \frac{\Phi_0}{2\pi}I\frac{L_1}{L}\gamma_2 + \left(\frac{\Phi_0}{2\pi}\right)^2\left(\frac{1}{2L}\right)\left(\gamma_1 - \gamma_2 - 2\pi\frac{\Phi_a}{\Phi_0}\right)^2 \quad (2)$$

where $E_{ji} = I_{0i}\Phi_0/2\pi$ is the Josephson energy for the $i$-th junction, and $I_{0i}$ is the critical current of the i-th junction. Note that in this expression, we have assumed for simplicity that the mutual inductance between the arms of the SQUID can be neglected, so that the total inductance of the SQUID loop is $L = L_1+L_2$ [12]. The first two terms in the Lagrangian are just the energy stored in the two junction capacitances. To see this, note that the ac Josephson relations give:

$$V_1 = \frac{\Phi_0}{2\pi}\dot\gamma_1, \quad (3)$$

$$V_2 = \frac{\Phi_0}{2\pi}\dot\gamma_2, \quad (4)$$

where and $V_1$ and $V_2$ are the voltages across junctions $J1$ and $J2$ respectively.

Given the Lagrangian, we can now use Euler's equation to find the equation of motion for $\gamma_1$:

$$0 = \frac{\partial}{\partial t}\frac{\partial\mathcal{L}}{\partial\dot\gamma_1} - \frac{\partial\mathcal{L}}{\partial\gamma_1} = m_1\ddot\gamma_1 + \frac{\partial}{\partial\gamma_1}U(\gamma_1,\gamma_2) \quad (5)$$

Substituting Eq. (2) for $U$ and using the definition of $m_1$, we can write this in the form:

$$I\frac{L_2}{L} - \left(\frac{\Phi_0}{2\pi L}\right)\left(\gamma_1 - \gamma_2 - 2\pi\frac{\Phi_a}{\Phi_0}\right) = I_{01}\sin\gamma_1 + C_1\frac{\Phi_0}{2\pi}\ddot\gamma_1 \quad (6)$$

The left hand side of Eq. (6) can be simplified by using the flux-phase relation for the SQUID loop [12]:

$$\gamma_1 = \gamma_2 + \frac{2\pi}{\Phi_0}(-L_1 I_1 + L_2 I_2 + \Phi_a). \tag{7}$$

where $I_1$ and $I_2$ are the currents flowing through the $J1$ and $J2$ arms of the SQUID loop, respectively (see Fig. 2). From current conservation, the bias current must divide between the two arms of the SQUID, so $I=I_1+I_2$. Using this and the flux phase relation, we find:

$$I_1 = I\frac{L_2}{L} - \left(\frac{\Phi_0}{2\pi L}\right)\left(\gamma_1 - \gamma_2 - 2\pi\frac{\Phi_a}{\Phi_0}\right) \tag{8}$$

and Eq. (6) can then be put in the simple form:

$$I_1 = I_{01}\sin\gamma_1 + C_1\frac{\Phi_0}{2\pi}\ddot{\gamma}_1 \tag{9}$$

This is just what one would expect when current conservation is applied to the $J1$ arm of the SQUID; *i.e.* the current in the $J1$ arm of the SQUID is the sum of the supercurrent current through junction $J1$ (from the dc Josephson relation this is just $I_{01}\sin\gamma_1$) and the displacement current $C_1\dot{V}_1$ through the capacitor $C_1$.

Similarly, using Euler's equation we can find the equation of motion for $\gamma_2$:

$$I\frac{L_1}{L} + \left(\frac{\Phi_0}{2\pi L}\right)\left(\gamma_1 - \gamma_2 - 2\pi\frac{\Phi_a}{\Phi_0}\right) = I_{02}\sin\gamma_2 + C_2\frac{\Phi_0}{2\pi}\ddot{\gamma}_2 \tag{10}$$

Again applying $I=I_1+I_2$ and the flux-phase relation, we find:

$$I_2 = I\frac{L_1}{L} + \left(\frac{\Phi_0}{2\pi L}\right)\left(\gamma_1 - \gamma_2 - 2\pi\frac{\Phi_a}{\Phi_0}\right) \tag{11}$$

and we can write Eq. (10) as

$$I_2 = I_{02} \sin \gamma_2 + C_2 \frac{\Phi_0}{2\pi} \ddot{\gamma}_2 \tag{12}$$

Again, this is just what one would expect when current conservation is applied to the *J2* arm of the SQUID. Thus, the Lagrangian given by Eq. (1) yields equations of motion (6) and (10), and these are just the expected coupled equations of motion for the junction phases $\gamma_1$ and $\gamma_2$ in a dc SQUID.

The Hamiltonian of the dc SQUID can now be written as

$$H = p_1 \dot{\gamma}_1 + p_2 \dot{\gamma}_2 - \mathcal{L} = \frac{p_1^2}{2m_1} + \frac{p_2^2}{2m_2} + U(\gamma_1, \gamma_2), \tag{13}$$

where the canonical momenta are found from $p_i = \frac{\partial \mathcal{L}}{\partial \dot{\gamma}_i} = m_i \dot{\gamma}_i$. Using the standard prescription of quantum mechanics, we identify $p_i = -i\hbar \partial / \partial \gamma_i$ and $\gamma_i$ as operators.

To proceed, we now assume that the external flux $\Phi_a$ and the bias current $I$ are changed simultaneously such that $\Phi_a = L_1 I$. This simultaneous ramping of the applied flux and the current is done in experiments so that the static applied current mainly flows through *J1* while leaving *J2* unbiased, as discussed above. In this case, the Josephson inductance [19] of the isolation junction *J2* is a minimum and one obtains the highest isolation of the qubit junction *J1* from current noise in the bias leads. With this assumption the terms in the potential *U* that are proportional to just $\gamma_2$ cancel and we find,

$$U(\gamma_1, \gamma_2) = -E_{J1} \cos \gamma_1 - E_{J1} \frac{I}{I_{01}} \gamma_1 - E_{J2} \cos \gamma_2 + E_L (\gamma_1 - \gamma_2)^2, \tag{14}$$

where $E_L = (\Phi_0/2\pi)^2/2L$ sets the scale for the inductive coupling energy between the two junctions and we have ignored an offset term that does not depend on $\gamma_1$ and $\gamma_2$. Of course if there are fluctuations in the current I that are independent of fluctuations in $\Phi_a$, then Eq. (14) would not be appropriate. We consider this situation below in Section VI.

Figure 2 shows different views of the two-dimensional potential $U$ given by Eq. (14). The parameters for the simulation are listed in Table 1, and they correspond to typical values used in experiments. In particular, the device is biased with $I/I_{01} = 0.95$ and simultaneously biased with external flux as described above. For the region of $\gamma_1$ and $\gamma_2$ shown in the figure the potential is dominated by a washboard potential along $\gamma_1$ [see Figs. 2(c) and 2(d)], just as in the ideal phase qubit and a parabolic potential along $\gamma_2$ [see Fig. 2(b)]. The depths of the local minima are very different along the two directions. For these conditions, wells are 25 times shallower along the $\gamma_1$ direction than along the $\gamma_2$ direction. Along $\gamma_1$ the potential near a local minimum behaves like a harmonic potential with a significant additional cubic term, while along $\gamma_2$ it acts as a simple harmonic potential with a much smaller anharmonic component.

The potential $U$ also contains coupling between $\gamma_1$ and $\gamma_2$ that is proportional to $E_L$. Since the overall energy scale of the Hamiltonian is set by $E_{j1}$ and $E_{j2}$, we can construct a dimensionless coupling constant [20]:

$$\kappa_0 = \frac{E_L}{(E_{j1} + E_{j2})/2} = \frac{1}{2\pi\beta} \tag{15}$$

where $\beta = L(I_{01} + I_{02})/\Phi_0$ is the SQUID modulation parameter. Thus we see that increasing the loop inductance L and critical currents of the junctions causes the coupling between the junctions to decrease and that the weakest coupling is achieved for $\beta \gg 1$. In this limit, if the critical currents are not too different, the loop will be able to trap a persistent circulating current and a corresponding trapped flux. There will be a maximum number of such persistent current states equal to the number of wells in the potential $U$, depending on L and the smaller of the two critical currents, with the trapped flux in different states differing by approximately $n\Phi_o$, where $n$ is an integer. We note that for the best isolation from the bias leads, the device would be run with no circulating current, corresponding to $n = 0$ in Fig 2(b) and no current through the isolation junction $J2$. High flux states, up to $n = \pm 8$ in the case of Fig. 2(b), correspond to loading $J2$ with current which increases its Josephson inductance and reduces the isolation.

For phase qubits we are mainly interested in the limit $E_{j1} \sim E_{j2} \gg e^2/2C_1 \sim e^2/2C_2$. In this limit the junction phases are typically well-localized in a given well. Experimentally, the system can be initialized in any of the local minima [20, 21] and the behavior depends on which well is chosen. The position and depth of each minimum depends on the bias current. For a particular well, the minimum disappears at a critical bias current $I_c$. In general, this bias critical current is not equal to either the critical current $I_{01}$ or $I_{02}$ of the individual junctions. In Fig. 3(a) we show the critical bias current $I_c$ for different wells along the $\gamma_2$ direction, while remaining in the well closest to $\gamma_1 = 0$. It is the value of $I_c$ where a particular minimum of the potential $U$ reduces to a saddle point and is calculated numerically. The nearly linear dependence of the critical current with well number is a consequence of the large loop inductance, which gives an effective small coupling between $\gamma_1$ and $\gamma_2$; the different minima correspond to circulating currents in the SQUID loop that differ by about $\Phi_0/L$.

In Fig 3(b) we show the plasma frequency $\omega_2$ of the isolation junction as a function of well number along $\gamma_2$, where $m_2\omega_2^2 = d^2U(\gamma_1,\gamma_2)/d\gamma_2^2$ at the minima. In the absence of the small coupling term $E_L(\gamma_1-\gamma_2)^2$, the minima along $\gamma_2$ would be spaced by $2\pi$ and the plasma frequency $\omega_2$ would be independent of the well index. The coupling term causes the distance between the minima to decrease, producing a quadratic decrease of $\omega_2$ with increase in well index number.

### III. CUBIC APPROXIMATION

It is not possible to obtain exact analytical solutions to the Schrödinger's equation for the Hamiltonian given by Eq. (13). However, since phase qubits are in the limit $E_{j1} \sim E_{j2} >> e^2/2C_1 \sim e^2/2C_2$, the junction phases can be relatively well-localized in a given well. Useful approximate results can then be found by making an expansion of the potential near a local minimum $(\gamma_1^m, \gamma_2^m)$ of the well and truncating at the cubic terms. For $I$ near the critical current $I_c$ of this well, the Hamiltonian can be approximated as

$$\overline{H} \cong \frac{p_1^2}{2m_1} + \frac{1}{2}m_1\omega_1^2(\gamma_1-\gamma_1^m)^2 - g_1(\gamma_1-\gamma_1^m)^3 + \frac{p_2^2}{2m_2} + \frac{1}{2}m_2\omega_2^2(\gamma_2-\gamma_2^m)^2 \qquad (16)$$
$$+ g_{12}(\gamma_1-\gamma_1^m)(\gamma_2-\gamma_2^m) + U(\gamma_1^m,\gamma_2^m),$$

where

$$m_1\omega_1^2 = E_{J1}\cos\gamma_1^m + 2E_L$$
$$\approx E_{J1}\left(1-\left(\frac{I-I_c}{I_{01}}+\sin\gamma_1^*\right)^2\right)^{1/2} + 2E_L \qquad (17)$$

$$m_2 \omega_2^2 = E_{J2} \cos \gamma_2^m + 2E_L \,, \tag{18}$$

$$g_1 = \frac{1}{6} E_{J1} \sin \gamma_1^* \,, \tag{19}$$

$$g_{12} = -2E_L \,, \tag{20}$$

$$\gamma_1^* = \arccos(-2E_L / E_{J1}), \tag{21}$$

$$\gamma_2^m \approx 2\pi k + 2E_L (\gamma_1^* - 2\pi k)/ E_{J2}, \tag{22}$$

$$I_c = I_{01} \sin \gamma_1^* + \frac{2E_L}{\Phi_0 /(2\pi)} (\gamma_1^* - \gamma_2^m). \tag{23}$$

where the integer $k$ is the well index, and $I_c$ is the critical bias current. The cubic term in $(\gamma_1 - \gamma_1^m)$ is proportional to $g_1$ while the coupling between $\gamma_1$ and $\gamma_2$ is proportional to $g_{12}$. Since $E_L \ll E_{J1}$, Eq. (21) implies under typical bias condition that $\gamma_1^*$, which is the critical value of $\gamma_1$ at which the washboard minima turns into a saddle point, is close to $\pi/2$ and $\sin \gamma_1^* \approx 1$. We note that in this truncated Hamiltonian, the $(\gamma_1 - \gamma_2)$ coupling term has been retained exactly, and the higher order terms in $(\gamma_1 - \gamma_1^m)$ and $(\gamma_2 - \gamma_2^m)$ that have been dropped are about 3 orders of magnitude smaller than the coupling term.

Equations (16-23) reproduce the nearly linear dependence of the numerically-determined critical current with well index along $\gamma_2$, as seen in Fig. 3(a), as well as the quadratic dependence of the plasma frequency shown in Fig. 3(b). Deviations from the quadratic behavior, however, are significant for larger well-number index. Also, the expression for $\omega_1$ given in Eq. (17) has some similarities to the single qubit case [11], but in fact has a different form and differs quantitatively.

The eigenvalues and eigenfunctions of the Hamiltonian $\overline{H}$ can be found by treating the cubic and linear coupling terms as small perturbations to the rest of the Hamiltonian, and then applying second-order time-independent perturbation theory. Tunneling rates can then be determined using the WKB (Wentzel-Kramers-Brillouin) approximation. Details and analytical results from this analysis are summarized in Appendix A.

## IV. NUMERICAL SIMULATIONS

For comparison, we also numerically solved Schrödinger's equation for the Hamiltonian $H$ given by Eqs. (13-14) and obtained the energy levels, tunneling rates, and the wave function of the various resonant states. For this numerical simulation, we used the method of complex scaling [22]. In this method we make the substitutions $\gamma \to \gamma e^{i\theta}$ and $p_\gamma \to p_\gamma e^{-i\theta}$ for both $\gamma_1$ and $\gamma_2$, using the same angle $\theta$ in the Hamiltonian, and find the complex eigenvalues and eigenvectors of the resulting non-Hermitian Hamiltonian. This transformation allows Gamow–Siegert states (subject to outgoing boundary wave conditions) to be calculated in a finite basis set. The complex eigenvalues are of the form $E - i\hbar\Gamma/2$, where the real part is the energy of the metastable state and the imaginary part gives the tunneling rate $\Gamma$. To discretize the Hamiltonian we use a two dimensional harmonic oscillator basis localized about the relevant potential minimum with frequencies equal to the plasma frequencies along $\gamma_1$ and $\gamma_2$. We find convergence upto 0.1% is reached for a basis set of $30\times30$ harmonic oscillator states.

## V. COMPARISON BETWEEN NUMERICAL SIMULATIONS AND PERTURBATION RESULTS

The symbols in Figure 4(a) show calculated energy levels vs. bias current $I$ for the $k = 0$ well, as found from the numerical simulation using complex scaling. In this case, we have set the coupling between the junctions to $\kappa_0 = 0.0040$ (see Table 1 for the rest of the device parameters) and assumed simultaneous flux and current ramping. For comparison, the solid curves show results for the energy levels when $\kappa_0 = 0$, calculated from the analytical results for single junctions (see Appendix A). As expected since the numerical simulations were done for quite weak coupling, the results from these two calculations are virtually identical on this scale.

In Fig. 4, the bias current runs from $I = 0.90 I_c$ to $0.97 I_c$, i.e. close to the critical current of 17.82 $\mu$A of the $k = 0$ well. These currents correspond to reasonable conditions for reading out the state by a tunneling measurement, such as the experiments in ref. [2]. We note that this critical current exceeds $I_{01} = 17.75$ $\mu$A because some of the applied current is diverted through junction J2; even with a simultaneous flux and current ramp, the change in Josephson inductance with current causes some imbalance. The eigenstates are labeled by state $|n,m\rangle$ where $n$ is the excitation or energy level in the qubit junction and $m$ is the excitation or energy level in the isolation junction (see Appendix A). In Fig. 4(a) only the states $|00\rangle$ and $|10\rangle$ lie below the top of the saddle point of the two-dimensional potential, which might suggest that only these state have a lifetime that is long compared to the oscillation period of the harmonic trap. This, however, is not true. States with one excitation in the isolation junction, i.e. $|01\rangle$ and $|11\rangle$, are also long lived against tunneling. This is because the isolation junction is well described by a harmonic potential and the coupling between the two junctions is small.

The state $|20\rangle$ also lives relatively long, but only because it lies just slightly above the barrier.

Figure 4(a) also shows two examples of crossings in the energy level diagram. These are actually avoided crossings, although the splitting is too small to be seen on this scale. At each crossing, an excited state of the qubit junction ($n = 3$ or 4) is in resonance with a state with one excitation ($m = 1$) in the isolation junction. Since the coupling constant $\kappa_0$ is small, the mixing is small and the resulting splittings are very small (approximately 6 MHz). As a result, in an experiment where the current is being swept, it will be relatively easy to be diabatic with respect to the crossings.

The symbols in Fig. 4(b) show the corresponding tunneling rate versus current for the different states, as found from the numerical simulations using complex scaling. Again, the solid curves show results when the coupling $\kappa_0 = 0$, found using the analytical results for single junctions given in Appendix A. We note that there is no discernible difference between the coupled and uncoupled results, except for states |01> and |11> where the isolation junction is excited. For these excited states of the isolation junction, the perturbative effects of the coupling term are needed because the tunneling rates are sensitive to even small mixings with highly-excited levels of the qubit junction. Here the $|01\rangle$ state mixes with $|30\rangle$ while the $|11\rangle$ state mixes with $|40\rangle$ [see Fig. 4(a)]. For comparison, the dashed curves show the results when the coupling term is included using second order perturbation theory; we find excellent agreement with complex scaling. From these results, we conclude that as long as we stay away from the very narrow avoided crossings, the dc SQUID phase qubit behaves much like a single phase qubit and the coupling can be safely ignored.

Figure 5 shows energy levels and tunneling rates for well $k = 8$ along the $\gamma_2$ direction. The critical current is reduced from the $k=0$ value of 17.82 µA to just 13.02 µA. As shown in Fig. 3 the plasma frequency along the $\gamma_2$ direction decreases with well number, hence, avoided crossing between energy levels occur at smaller energies. In fact, the $|01\rangle$ state now avoids the $|20\rangle$ state and the $|11\rangle$ avoids $|30\rangle$. The effects of coupling are stronger than they were for $k = 0$ and this is reflected in the more gradual changes in the tunneling rates with current $I$ near the avoided crossings. Consequently, as well number $k$ increases, the circuit behaves less and less like an ideal phase qubit.

## VI. RESPONSE OF THE SYSTEM TO LOW FREQUENCY NOISE

Low frequency current noise in the bias current $I$ and flux current $I_f$ can be a significant source of spectral resonance broadening and decoherence in phase qubits [2, 10, 11]. In addition, the bias leads present a dissipative impedance to the junction that decreases the lifetime of excited states. The main idea behind the design of the dc SQUID phase qubit is to use $L_1$ and $L_2$ as an inductive isolation network that steps up the lead impedance and prevent current noise from reaching the qubit. Here we examine how well this design works and what effects it produces on the qubit junction.

To understand how well the qubit junction is isolated from current noise in the bias current $I$, we first consider the classical description. When a small change $\Delta I$ is made in the bias current $I$, the external flux is held fixed at $L_1 I$. Provided the fluctuations are slow enough that the displacement current in the junction capacitors can be neglected, the change $\Delta I_1$ in the current through the qubit junction can be found by treating the

circuit as an inductive current divider. We find:

$$\Delta I_1 = (L_2 + L_{J2})\Delta I /(L + L_{J1} + L_{J2}). \tag{24}$$

where $L_{Ji} = \Phi_0 / 2\pi I_{0i} \cos \gamma_i^m$ is the Josephson inductance of i-th junction. From Eq. (24), it is useful to define the current noise power isolation factor:

$$r_I = \left(\frac{\Delta I}{\Delta I_1}\right)^2 = \left(\frac{L + L_{J1} + L_{J2}}{L_2 + L_{J2}}\right)^2 \tag{25}$$

For large $r_I$, the qubit junction is well-isolated from current noise in the leads. For the parameters shown in Table 1, $L + L_{j1} + L_{j2} \sim 3.4$ n H $\gg L_2 + L_{j2} \sim 40$ pH, so that $r_I \sim 7000$, and the network is quite effective at filtering out current fluctuations.

To understand the effect of the network on dissipation, we now assume that the current bias leads have a real impedance $Z_0 \sim 50\ \Omega$. The network steps up this impedance so that the effective parallel impedance $Z_{eff}$ that shunts the qubit junction is found from:

$$\frac{1}{Z_{eff}} = \frac{1}{R_{eff}} + \frac{1}{i\omega L_{eff}}. \tag{26}$$

Here $R_{eff}$ is the effective shunting resistance:

$$R_{eff} = \left(\frac{L_1 + L_2 + L_{J2}}{L_2 + L_{J2}}\right)^2 Z_o + \frac{\omega^2 L_1^2}{Z_o} \cong \left(\frac{L_1 + L_2 + L_{J2}}{L_2 + L_{J2}}\right)^2 Z_o = r_I Z_o \tag{27}$$

and $L_{eff}$ is the effective parallel shunting inductance:

$$L_{eff} = \frac{(L_1 + L_2 + L_{J2})^2 + \left(\frac{\omega(L_1 + L_2 + L_{J2})}{Z_o}\right)^2}{L_1 + L_2 + L_{J2} + \left(\frac{\omega(L_2 + L_{J2})}{Z_o}\right)^2 L_1} \cong L_1 + L_2 + L_{J2}. \tag{28}$$

For typical parameters and the low frequencies of interest, the frequency dependent terms can be neglected, as indicated. The lifetime of the excited state of the qubit then becomes:

$$T_1 = \left(\frac{1}{R_{eff}} + \frac{1}{R_1}\right)^{-1} C_1$$

where $R_1$ accounts for any resistance shunting the junction that is from sources other than the leads.

Finally, if we ignore the inductive network, the frequency at which the phase of the qubit junction executes small oscillations is given by [2]

$$\omega_1 = \frac{1}{\sqrt{L_{j1}C_1}} = \omega_{p1}\left(1 - \left(\frac{I_1}{I_{01}}\right)^2\right)^{1/4}, \qquad (29)$$

where $I_1$ is the current flowing through the qubit junction and

$$\omega_{p1} = \sqrt{\frac{8E_{C1}E_{J1}}{\hbar^2}} = \sqrt{\frac{2\pi I_{01}}{C_1 \Phi_0}}. \qquad (30)$$

is the plasma frequency of the isolated junction when it is not biased. Including the rest of the circuit produces a small shift in the plasma frequency due to the shunting inductance $L_{eff}$, and we find a perturbed resonance frequency $\omega'$ given by:

$$\omega_1' = \sqrt{\left(\frac{1}{L_{j1}} + \frac{1}{L_{eff}}\right)\frac{1}{C_1}} \approx \omega_1 + \frac{1}{2\omega_1(L_1 + L_2 + L_{J2})C_1}. \qquad (31)$$

Given a change $\Delta I$ in the bias current, the change in the plasma frequency is then:

$$\Delta\omega_1' = \frac{d\omega_1'}{dI_1}\frac{(L_2 + L_{J2})}{(L + L_{J1} + L_{J2})}\Delta I \approx -\frac{1}{2}\frac{(I/I_c)^2}{\left(1 - (I/I_c)^2\right)^{3/4}}\frac{L_2 + L_{J2}}{L + L_{J2}}\frac{\Delta I}{I} \qquad (32)$$

where in the last approximation we neglected the contribution of the small correction term in Eq. (31). We can now define the classical response function $G_c$ through the relationship:

$$\Delta\omega_1' = \omega_1' \frac{\Delta I}{I} G_c$$

which is the ratio between the relative change in the plasma frequency to the relative change in the bias current. From Eq. (32) we find:

$$G_c \approx -\frac{1}{2}\frac{(I/I_c)^2}{\left(1-(I/I_c)^2\right)^{3/4}}\frac{L_2+L_{J2}}{L+L_{J1}+L_{J2}} \tag{33}$$

and should be of the order of 1 or less for good isolation.

To perform a quantum analysis of this situation, we now assume that the fluctuations $\Delta I$ are slow enough that time-independent perturbation theory can be used. We are interested in the situation where the current noise is not accompanied by a simultaneous fluctuation in the external flux. In this case, the Hamiltonian picks up an additional term:

$$\Delta H = -\frac{\Phi_0}{2\pi}\Delta I(t)\left(\frac{L_2}{L}\gamma_1 + \frac{L_1}{L}\gamma_2\right). \tag{34}$$

As we have seen in Sec II, the SQUID's Hamiltonian is well approximated by a cubic potential along $\gamma_1$ and a simple harmonic potential along $\gamma_2$ with a small coupling proportional to $\kappa_0$. With this Hamiltonian and the matrix elements listed in Appendix A, we can find analytical expressions for level shifts due to the perturbation $\Delta H$. The corrections to the energy of the ground and first excited state of the qubit up to second order are

$$\Delta_{n0} = -\frac{\Phi_0}{2\pi}\Delta I(t)\left(\frac{L_2}{L}\langle n0|\gamma_1-\gamma_1^m|n0\rangle + 2\frac{L_1}{L}\langle n1|\gamma_2-\gamma_2^m|n0\rangle\frac{\langle n0|\kappa(\gamma_1-\gamma_1^m)(\gamma_2-\gamma_2^m)|n1\rangle}{E_{n0}-E_{n1}}\right) \tag{35}$$

for $n = 0$ and 1, respectively. In this equation we have only retained terms proportional to $\Delta I$. After some algebra we find,

$$\Delta\omega_1 = \frac{\Delta_{10} - \Delta_{00}}{\hbar} = -\left(\frac{\Phi_0}{2\pi\hbar}\right)\frac{L_2 + L_{J2}}{L + L_{J2}}\left(\frac{1}{2}\left(\frac{2\pi}{\Phi_0}\right)^2 \frac{I_{01}\sin\gamma_1^* \sqrt{\hbar}}{C_1^{3/2}\omega_1^{5/2}}\right)\Delta I$$

$$\approx -\left(\frac{\Phi_0}{2\pi\hbar}\right)\frac{L_2 + L_{J2}}{L + L_{J2}}\left(\frac{1}{2}\left(\frac{2\pi}{\Phi_0}\right)^2 \frac{I_{01}\sqrt{\hbar}}{C_1^{3/2}\omega_1^{5/2}}\right)\Delta I. \quad (36)$$

Analogous to the classical response function $G_c$, we can now define the quantum mechanical response function $G_q$ through $\Delta\omega_1 = \frac{(E_{10} - E_{00})}{\hbar}G_q\frac{\Delta I}{I}$. This equation should be compared to Eq. 33 from our classical analysis. Here the frequency $(E_{10} - E_{00})/\hbar$ plays the role of the classical plasma frequency $\omega_1$. Like the classical response function $G_c$, the quantum response function $G_q$ should be of the order of 1 or less for good isolation. In Fig. 6 we compare the classical and quantum mechanical response functions for well number 0 and 8 as a function of $I/I_c$. We see that the classical and quantum responses are nearly the same for currents $I/I_c < 0.97$. For smaller well index, the two curves agree over an even wider range of $I/I_c$. Both functions increase with current and are on the order of 1 for $I/I_c \approx 0.990$. Hence only for currents less than $0.990 I_c$ does the isolation scheme work well. Furthermore in higher wells, the isolation becomes worse for the same value of $I/I_c$. Thus, the best isolation is achieved in the lowest well number; this is just what is expected from the classical analysis since this minimizes the Josephson inductance of the isolation junction. It is worth noting that by preparing the system in different wells, we can control the sensitivity of the system to bias current noise, and this is of considerable practical value in determining whether noise is entering the system through the leads [23].

To understand the effect of noise in the flux current $I_f$ we note that the entire analysis above and in particular Eq. (36) is also valid for uncompensated low-frequency changes in the flux $\Delta\Phi_a$ if $\Delta I$ is replaced by $\Delta\Phi_a / L_2 = M\Delta I_f / L_2$. The typical experimental value of $M$ is given in Table 1 which is of the same order of magnitude as $L_2$. It is chosen to be sufficiently small to ensure that flux noise does not couple to the SQUID via this coil. For values we have chosen $\Delta\Phi_a / L_2 = 0.8\Delta I_f$. Thus the effect of flux noise is of the same order of magnitude as the current noise.

## VII. STATE DEPENDENT CURRENT REDISTRIBUTION

In this section we examine how the state of the qubit determines the current flowing through both arms of the SQUID. Figure 7 shows the expectation value of the current flowing through the isolation junction $\langle I_2 \rangle$ as a function of $I$ for well number 0 for four eigenstates of the Hamiltonian, i.e. $|00\rangle$, $|10\rangle$, $|01\rangle$ and $|11\rangle$. We find that the current changes on the order of a few nA when the qubit junction is in an excited state, while it is virtually independent of the excitation in the isolation junction. The current $\langle I_2 \rangle$ increases with bias current for all states. We have also studied the dependence of $\langle I_2 \rangle$ on the well number along the $\gamma_2$ direction. The results are qualitatively the same except that the actual current is offset or shifted by $\Phi_0/L \sim 1\,\mu A$ as we go from one well to the next.

As evident from fig 7, the difference in $\langle I_2 \rangle$ between the ground state and the first excited state persists even when I is very small, i.e. in a regime where $1/\Gamma \approx \infty$. So if we can measure this small change in $\langle I_2 \rangle$ which is of the order of few nA, in that regime, the measurement will be non - destructive in so far that the qubit remains in its well or within the Hilbert space. This is in stark contrast to current schemes of measurement in phase qubits by tunneling the qubit through the barrier thereby destroying it completely. Measurement of <I2> is non destructive but should not be confused with quantum non–demolition kind of measurements used in the context of quantum measurement theory where certain set of quantum states are not changed by the measurement device. For the parameters used for the simulation, a very sensitive detector would be needed; the current only changes by a few nA in going from the ground state to the excited state, and one typically would like to read out the state in a few ns. The natural choice would be to use a second SQUID to detect the flux generated by this current. Another way is to use the so called Josephson Bifurcation Amplifier [24] device to detect small changes in $\langle I_2 \rangle$.

**VIII. CONCLUSIONS**

In conclusion, we analyzed the behavior of a dc SQUID phase qubit in which one junction acts as a phase qubit and the rest of the device provides isolation from dissipation and noise in the bias leads. We found the two-dimensional Hamiltonian of the system and used numerical methods and a cubic approximation to solve Schrödinger's equation for the eigenstates, energy levels, and tunneling rates. Using these results, we found that the dc SQUID phase qubit is well isolated from low frequency bias current noise and behaves as a single phase qubit when the parameters and the bias current are chosen correctly. We also examined the state-dependent redistribution of current between the two arms of the SQUID and noted that in principle this can be used for non-destructive readout.

## APPENDIX A

This appendix summarizes perturbation theory results for the Hamiltonian $\overline{H}$:

$$\overline{H} \cong \frac{p_1^2}{2m_1} + \frac{1}{2}m_1\omega_1^2(\gamma_1 - \gamma_1^m)^2 - g_1(\gamma_1 - \gamma_1^m)^3 + \frac{p_2^2}{2m_2} + \frac{1}{2}m_2\omega_2^2(\gamma_2 - \gamma_2^m)^2 \quad \text{(A1)}$$
$$+ g_{12}(\gamma_1 - \gamma_1^m)(\gamma_2 - \gamma_2^m) + U(\gamma_1^m, \gamma_2^m),$$

as in Eq. (16), where for notational convenience we set $\gamma_1^m = \gamma_2^m = 0$. For a typical dc SQUID phase qubit, the $g_{12}$ term is a small perturbation to the rest of the Hamiltonian; this term couples the phases $\gamma_1$ and $\gamma_2$ together. If we neglect this term the Hamiltonian separates. The motion along the $\gamma_2$ direction is harmonic, whereas that along $\gamma_1$ also has a cubic term proportional to $g_1$. The eigenstates of this uncoupled Hamiltonian up to second order in $g_1$ are denoted by $|n,m\rangle = |n\rangle|m\rangle$ where $n$ is the excitation in the qubit junction and $m$ is the excitation in the isolation junction.

The corresponding eigenenergies of this uncoupled Hamiltonian up to second order in $g_1$ are

$$E_{n,m} = \left((n+1/2) - \frac{30}{8}\lambda^2\left[\left(n+\frac{1}{2}\right)^2 + \frac{7}{60}\right]\right)\hbar\omega_1 + (m+1/2)\hbar\omega_2 \quad \text{(A2)}$$

where $n, m = 0, 1,$ etc. and $\lambda = g_1/\left[(m\omega_1/\hbar)^{3/2}\hbar\omega_1\right]$. The tunneling rates of the metastable energy levels of the cubic Hamiltonian can be found using a WKB (Wentzel-Kramers-Brillouin) approximation, and one finds [25]:

$$\Gamma_{n,m} = \frac{\omega_1}{\sqrt{2\pi}n!}(432N_s)^{n+1/2}\exp\left(-\frac{36}{5}N_s\right), \quad \text{(A3)}$$

where $N_s = 1/(54\lambda^2)$ is approximately the number of harmonic oscillator states with energy below the barrier of the cubic potential. The tunneling rate is independent of the state in the $\gamma_2$ direction. The above expression can be improved upon when $N_S$ is less than 2. We find empirically that a more accurate expression is given by

$$\Gamma_{n,m} = \frac{\omega_1}{\sqrt{2\pi n!}} (432 N_s)^{n+1/2} \frac{1}{1 + A\exp(36 N_s/5)}. \tag{A4}$$

where the fitting parameter $A$ turns out to be independent of $n$ and $m$ and can be determined from the exact numerical tunneling rates. Thus, the complex eigenenergies of the Hamiltonian $\overline{H}$ with the coupling $\kappa$ set to zero are $E_{n,m} - i\hbar\Gamma_{n,m}/2$.

For non-zero $\kappa$, we can compute corrections to the uncoupled eigenenergies using second order perturbation theory. We find the perturbative energy shift is

$$\Delta(E_{mn} - i\hbar\Gamma_{mn}/2) = \sum_{n'm'} g_{12}^2 \frac{|\langle nm|\gamma_1\gamma_2|n'm'\rangle|^2}{E_{nm} - i\hbar\Gamma_{nm}/2 - E_{n'm'} + i\hbar\Gamma_{n'm'}/2}, \tag{A5}$$

where the sum excludes $n'm' = nm$. The matrix elements in Eq. A5 reduce to simple products such as $\langle n|\gamma_1|n'\rangle\langle m|\gamma_2|m'\rangle$, where the matrix elements of $\gamma_2$ are determined by the properties of harmonic oscillator solutions. Up to second order in $\lambda$, the matrix elements $\gamma_{nn'} = \langle n|\gamma_1|n'\rangle$ are given by $\gamma_{0,0} = 3\lambda/2$, $\gamma_{0,1} = 1/\sqrt{2} + 11\sqrt{2}\lambda^2/8$, $\gamma_{0,2} = -\lambda/\sqrt{2}$, $\gamma_{0,3} = 3\sqrt{3}\lambda^2/8$, $\gamma_{1,1} = 9\lambda/2$, $\gamma_{1,2} = 1 + 11\lambda^2/2$, and $\gamma_{2,2} = 15\lambda/2$, for example [25].

**Table 1.** dc SQUID phase qubit parameters used in the calculations.

| | | | |
|---|---|---|---|
| $I_{01}$ (μA) | 17.75 | $I_{02}$ (μA) | 6.40 |
| $C_1$ (pF) | 4.44 | $C_2$ (pF) | 2.22 |
| $L_1$ (nH) | 3.39 | $L_2$ (pH) | 20.0 |
| $E_{J1}/h$ (THz) | 8.82 | $E_{J2}/h$ (THz) | 3.17 |
| $E_{C1}/h$ (MHz) | 3.82 | $E_{C2}/h$ (MHz) | 8.73 |
| $\beta$ | 39.83 | $\kappa_0 = 1/2\pi\beta$ | 0.0040 |
| $f_{p1}$ (GHz) | 16.42 | $f_{p2}$ (GHz) | 14.88 |
| $E_L/h$ (GHz) | 24.22 | $g_{12}$ (GHz) | 24.22 |
| $r_I$ | 7182 | $M$ (pH) | 16.0 |

**Figure Captions**

**Fig. 1.** (a) Schematic of a phase qubit. (b) Tilted washboard potential of the phase qubit. (c) Schematic of dc SQUID phase qubit. The circuit contains two Josephson junctions $J1$ and $J2$, with junction capacitance $C_1$ and $C_2$, respectively. The inductances of the two arms of the SQUID are $L_1$ and $L_2$. The current source $I$ biases the two junctions, while an external flux $\Phi_a$ is provided by the current source $I_f$ through the mutual inductance $M$. The currents flowing through the right and left arms of the SQUID are $I_1$ and $I_2$ respectively.

**Fig. 2.** Potential $U$ for the dc SQUID phase qubit with the parameters in Table 1 and I = 17.00 µA. (a) 2D surface plot of the potential. (b) Cross-section of $U$ along $\gamma_2$ for $\gamma_1 = 0$. The numbers label the individual wells. (c) Cross-section of $U$ along $\gamma_1$ for $\gamma_2 = 0$. (d) Cross-section of $U$ along $\gamma_1$ for $\gamma_2 = 0$ showing a single well.

**Fig. 3.** (a) Critical current $I_c$ vs. well number along $\gamma_2$. (b) Plasma frequency $\omega_2/2\pi$ vs. well number along $\gamma_2$. For both graphs the well-numbers are from Fig. 2(b).

**Fig. 4** (a) Energy levels and (b) tunneling rates for different metastable states for well $k = 0$ as a function of bias current $I$. The zero of the vertical axis is at the bottom of the metastable well. Symbols are from numerical calculation of the full 2D Hamiltonian using complex scaling, while the solid curves show analytical calculation where coupling between the two junctions is ignored. Dashed curves in panel b) show results from second-order perturbation theory. States are labeled by the ket $|n,m\rangle$, where the first index represents the qubit junction state and the second index represents the isolation junction state.

**Fig 5** (a) Energy levels and (b) tunneling rates for various metastable states for well $k = 8$ as a function of bias current $I$. The meaning of the symbols *etc*. is as in Fig. 4.

**Fig. 6** Absolute value of the response functions $G_c$ and $G_q$ as a function of current for well number $k = 0$ and 8. Points are obtained by exact numerical calculation, solid lines are from perturbation theory (see Sec VI) and dashed line shows predictions from classical analysis.

**Fig. 7** Expectation value of the current $\langle I_2 \rangle$ flowing through the isolation junction as a function of $I$ for well number 0. The points are numerical calculations based on the exact 2D Hamiltonian whereas the solid lines correspond to an analytical calculation with Hamiltonian $\overline{H}$ without the coupling term (see Eq. A1).

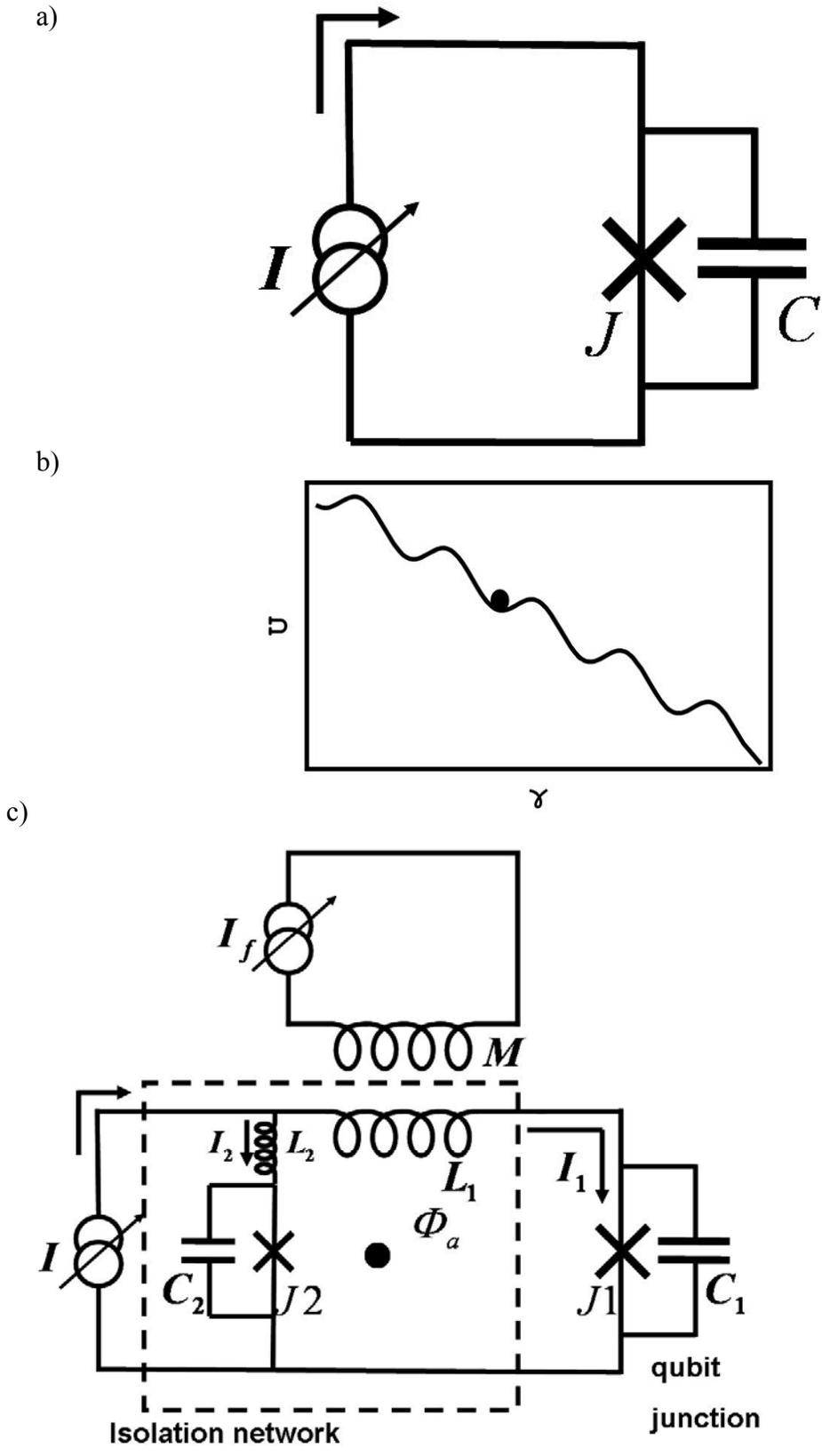

Fig 1, Mitra *et al*

a)
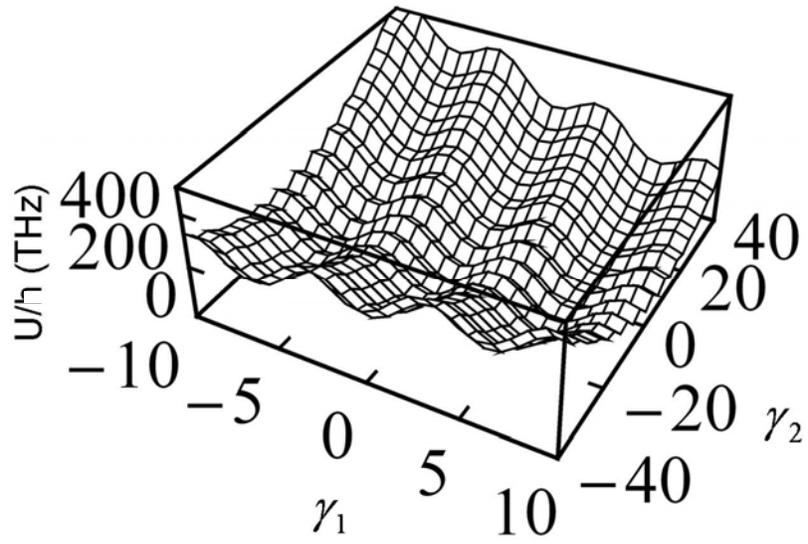

b)
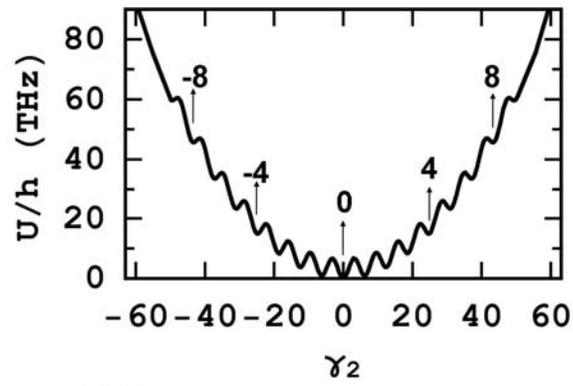

c)
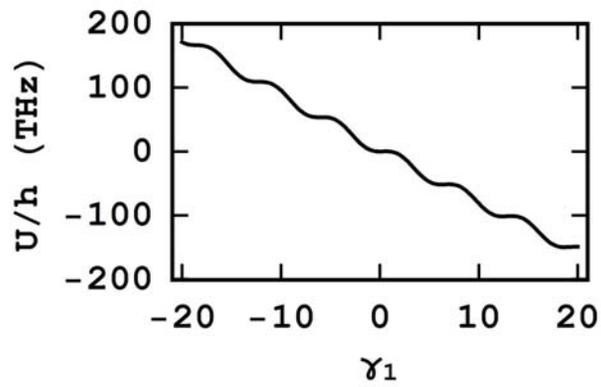

d)
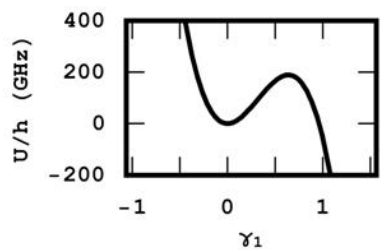

Fig 2, Mitra *et al*

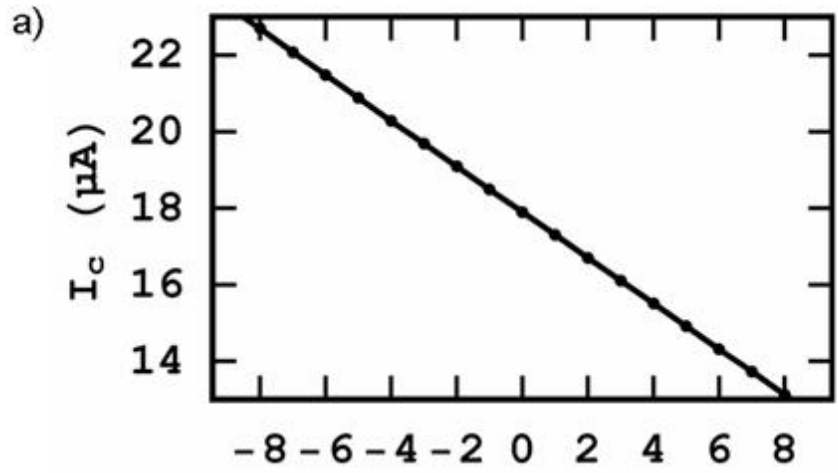
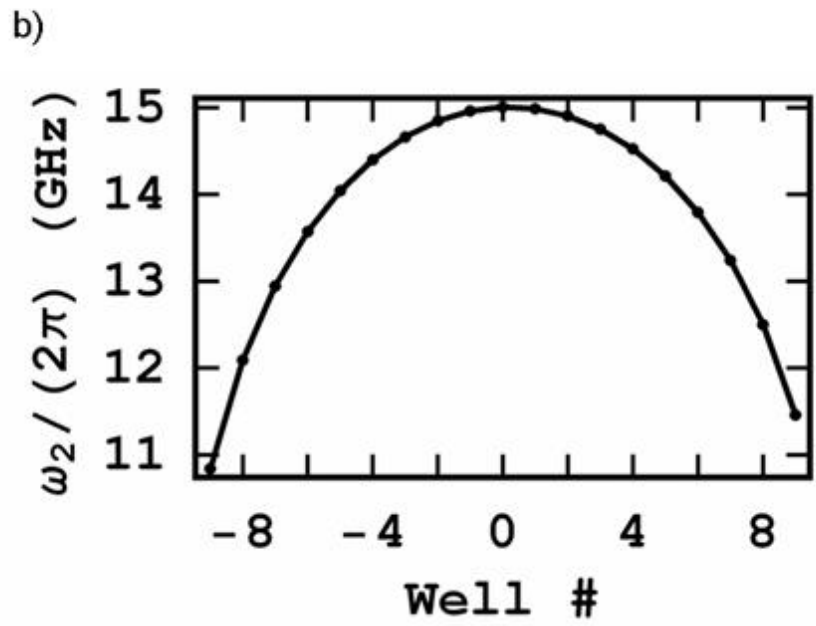

Fig 3, Mitra *et al*

a)

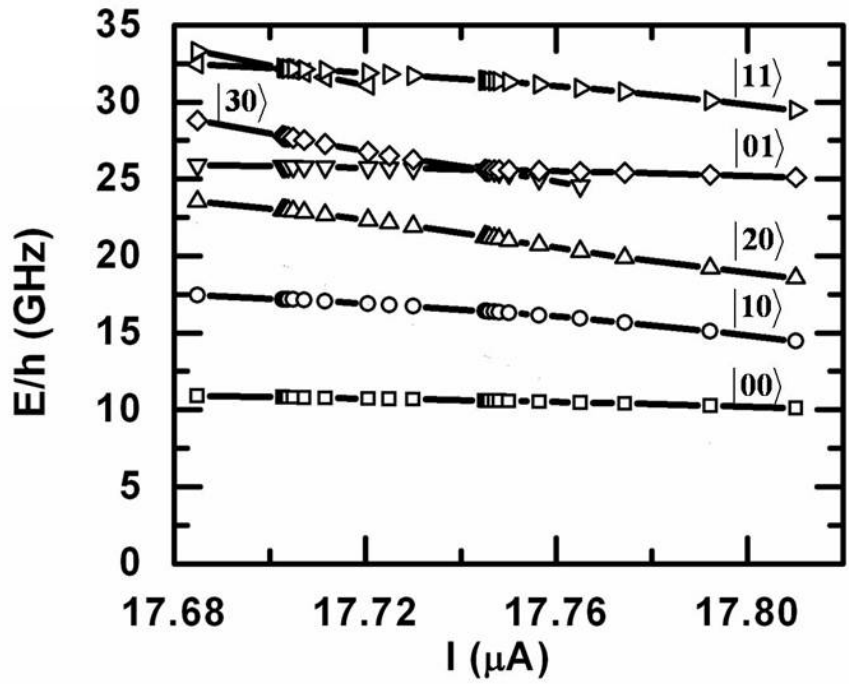

b)

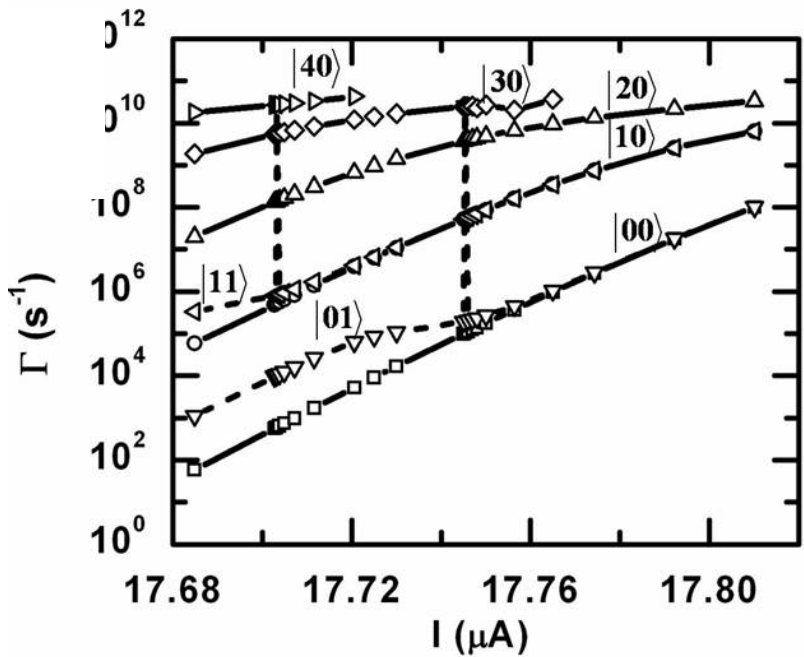

Fig 4, Mitra *et al*

a)

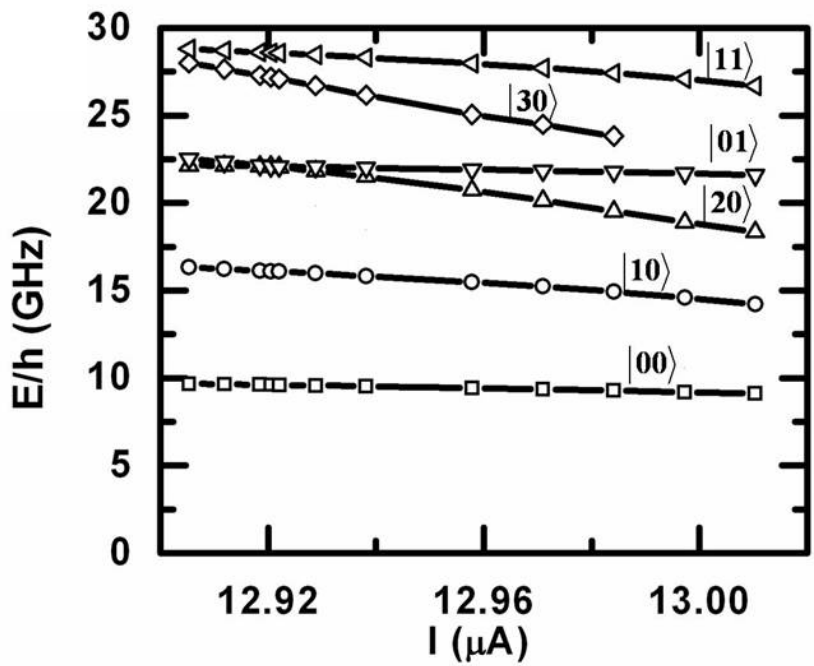

b)

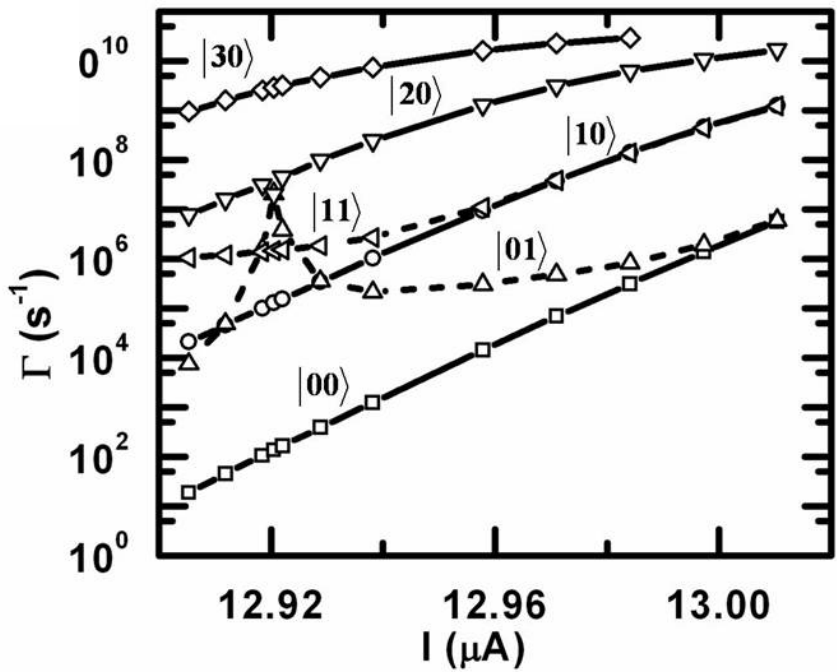

Fig 5, Mitra *et al*

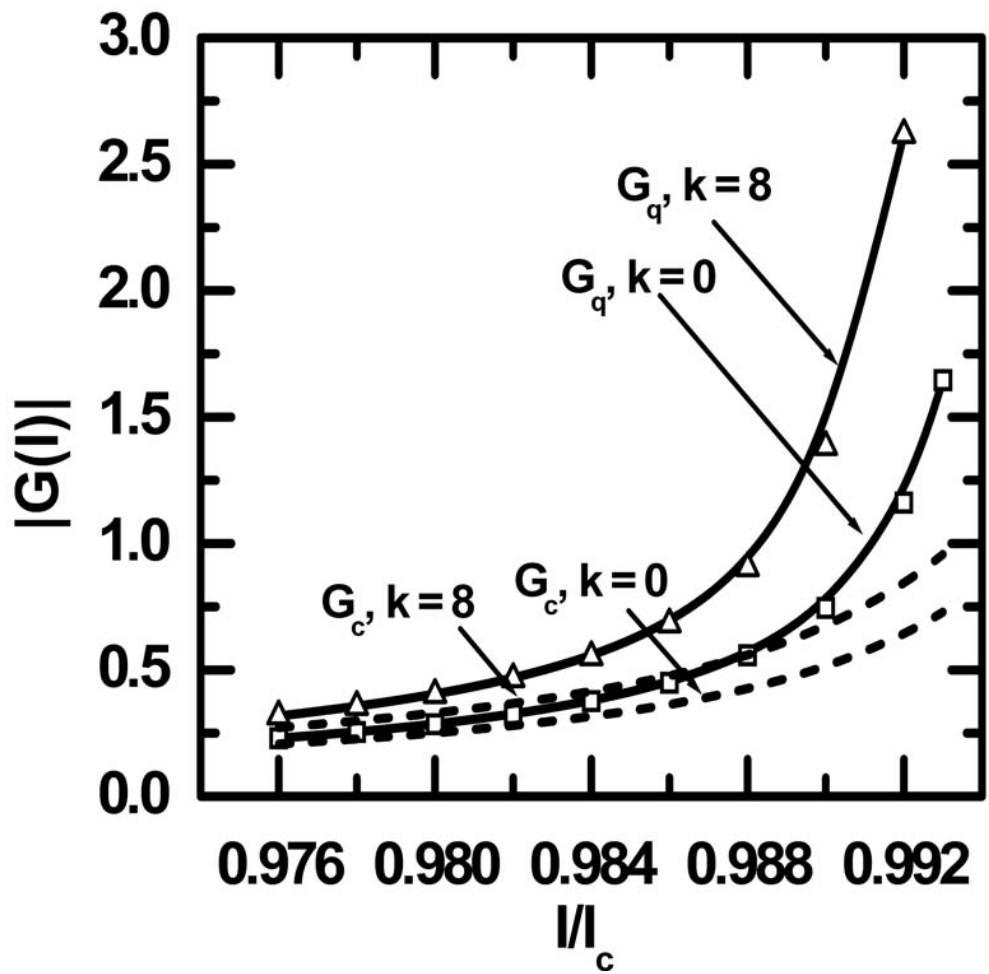

Fig 6, Mitra *et al*

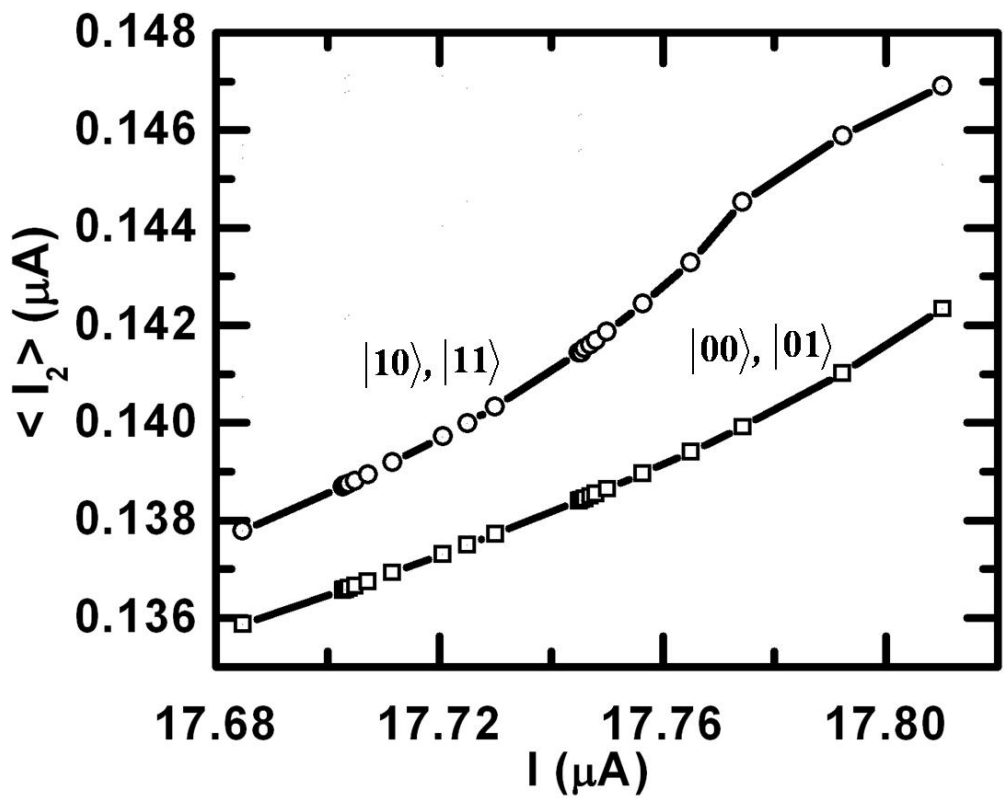